# 13,000-FOLD FREQUENCY DOWN CONVERSION FROM 280 MHZ TO 21 KHZ USING MECHANICAL FREQUENCY COMBS

*Ian Anderson, Zihuan Liu, Vivek Tallavajhula, Yen-Chen Wang, Tzu-Hsuan Hsu, Neal Hall, and Ruochen Lu*

The University of Texas at Austin, Tx, USA

## ABSTRACT

Optical frequency combs bridge the microwave and terahertz regimes by down-converting terahertz optical energy into microwave signals. A photodetector converts the comb's equally spaced lines, generated through optical nonlinearities, into a low-frequency tone at their spacing. Here, we present the first mechanical version of this process, using mechanical frequency combs to down-convert energy over 4 orders of magnitude, from the MHz to kHz regime. Thin-film lithium niobate (LN) is used to design an acoustic resonator with a wide range of mode frequencies. We achieve phononic comb excitation from 250 MHz to 900 MHz and compare the electrical and laser Doppler vibrometer (LDV) measurements at 277 MHz. We then show the low-frequency spectrum of the devices, highlighting mechanical motion at the comb's spacing and demonstrating energy transfer from 277 MHz to 21 kHz.

## INTRODUCTION

The ability to bridge between different signals of vastly different frequencies has revolutionized how we measure optical signals and established a consistent methodology for connecting radio frequency (RF) signals in the GHz regime to optical signals in the THz regime. This was first done via the use of the optical frequency comb, which ties the main frequency of the optical signal to that of the spacing of the modes in a resonator, typically a much lower frequency signal that is accessible via electronics [1]. These devices use optical nonlinearities to disperse energy into sideband tones, creating a structure of equally spaced lines in the frequency domain, which combine in the time domain as a pulsed signal with a lower repetition rate [2]. The signal is then fed to a photodetector, and the output is a clean signal with properties tied to the original optical signal. This process, known as optical frequency down-conversion, allows optical energy to be converted to RF frequencies, and finds uses in low-noise microwave generation [3], [4], and compact optical atomic clocks [5], [6].

Mechanical resonators, while much different from their optical or electromagnetic counterparts, follow fundamentally the same rules and processes: wave equation solutions in a cavity, with the same nonlinearities appearing at high powers. Thus, the search for the mechanical analogue of frequency combs has been conducted in parallel. First demonstrated in 2017 [7], these frequency combs function in a different way than integrated optical Kerr combs, as demonstrated in Fig. 1. Optical resonators use the third-order nonlinearity, termed the Kerr nonlinearity (hence their name), to distribute energy to their sidebands via four-wave mixing. In this process, nonlinearities from the optical signal counteract the dispersion of the optical signal to place the resonances at equal spacing from the input resonance and allow quality factor (Q) enhancement of the sideband signals. In mechanical resonators, achieving this same goal is much more difficult. Mechanical modes are highly dispersive, more spaced out, and nonlinearities are difficult to achieve in large mode numbers [8].

To this end, the work in [7] has demonstrated an alternative path to comb generation via second- and third-order nonlinearities. This method relies on parametric down conversion, whereby energy is first converted into two lower frequency tones in a three-wave mixing process. After this parametric down-conversion, further tuning gives rise to sidebands around each of the driven and parametric tones. The work has since been demonstrated in many other systems and materials, and a number of proposed applications [9], [10], [11]. These comb lines have spacings of much lower magnitude than those of the driven and parametric tones. The presence of these tones, along with the existence of frequency down-conversion in optics, leads us to look at the lower frequency spectrum of the mechanical comb devices to observe whether this frequency spacing coincides with low-frequency tones in the kHz regime. In fact, similar work has been demonstrated previously, showing 6:1 energy up-conversion through harmonics [12].

Our previous work has focused on higher frequency versions of these frequency combs, working with Lamb waves in the 100s of MHz in Lithium Niobate [13]. Here, we extend upon this work and present the first demonstration of mechanical frequency down-conversion over four orders of magnitude from these Lamb wave resonances to out-of-plane motion in the kHz regime via laser Doppler vibrometer (LDV) readout. This result extends the frequency comb down-conversion process beyond the optical domain and shows that a single compact mechanical device can link widely separated frequencies. The remainder of this paper is organized as follows. We first describe the design and fabrication of the device, then present the electrical and LDV measurements of the comb, and finally examine the low-frequency spectrum and its implications.

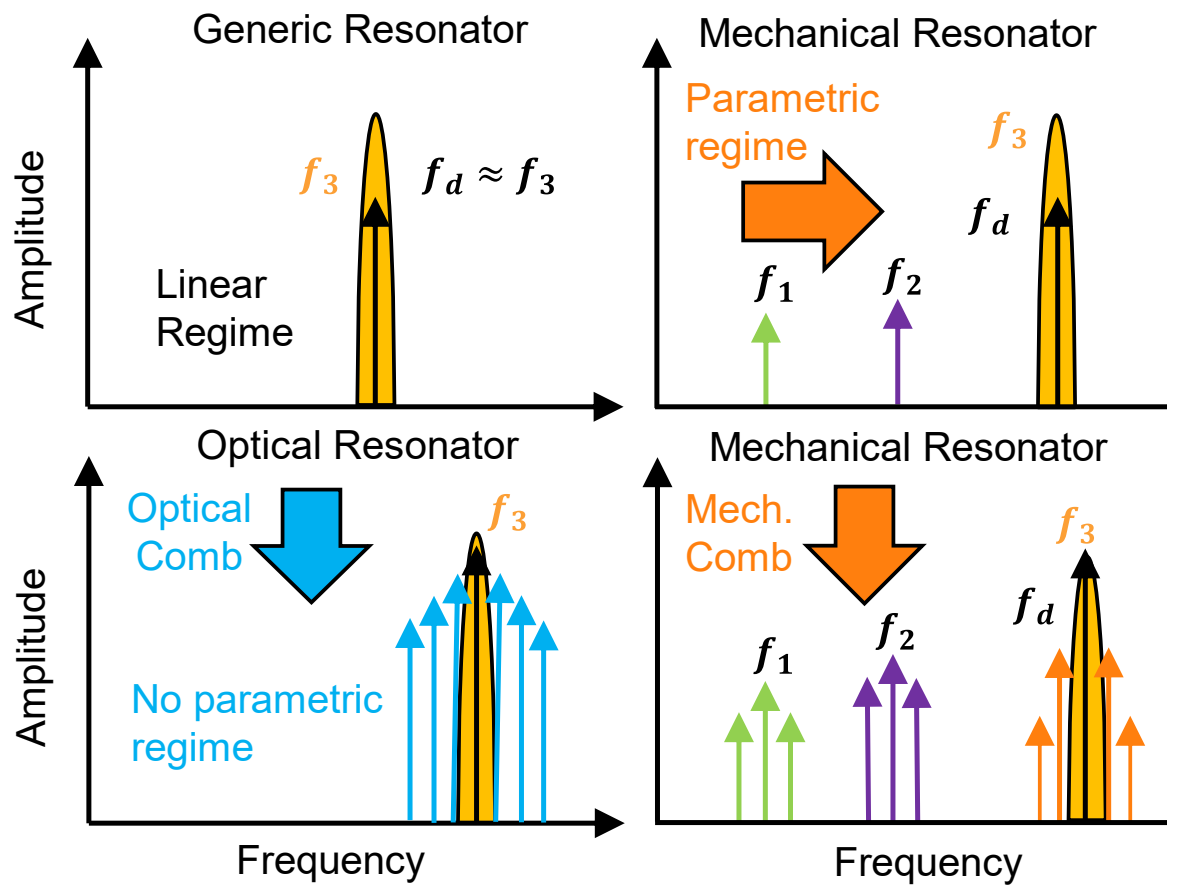

*Figure 1: (a) illustration of methodology for driving optical combs versus (b) methodology for driving of mechanical combs, showing pre-requisite parametric down conversion.*

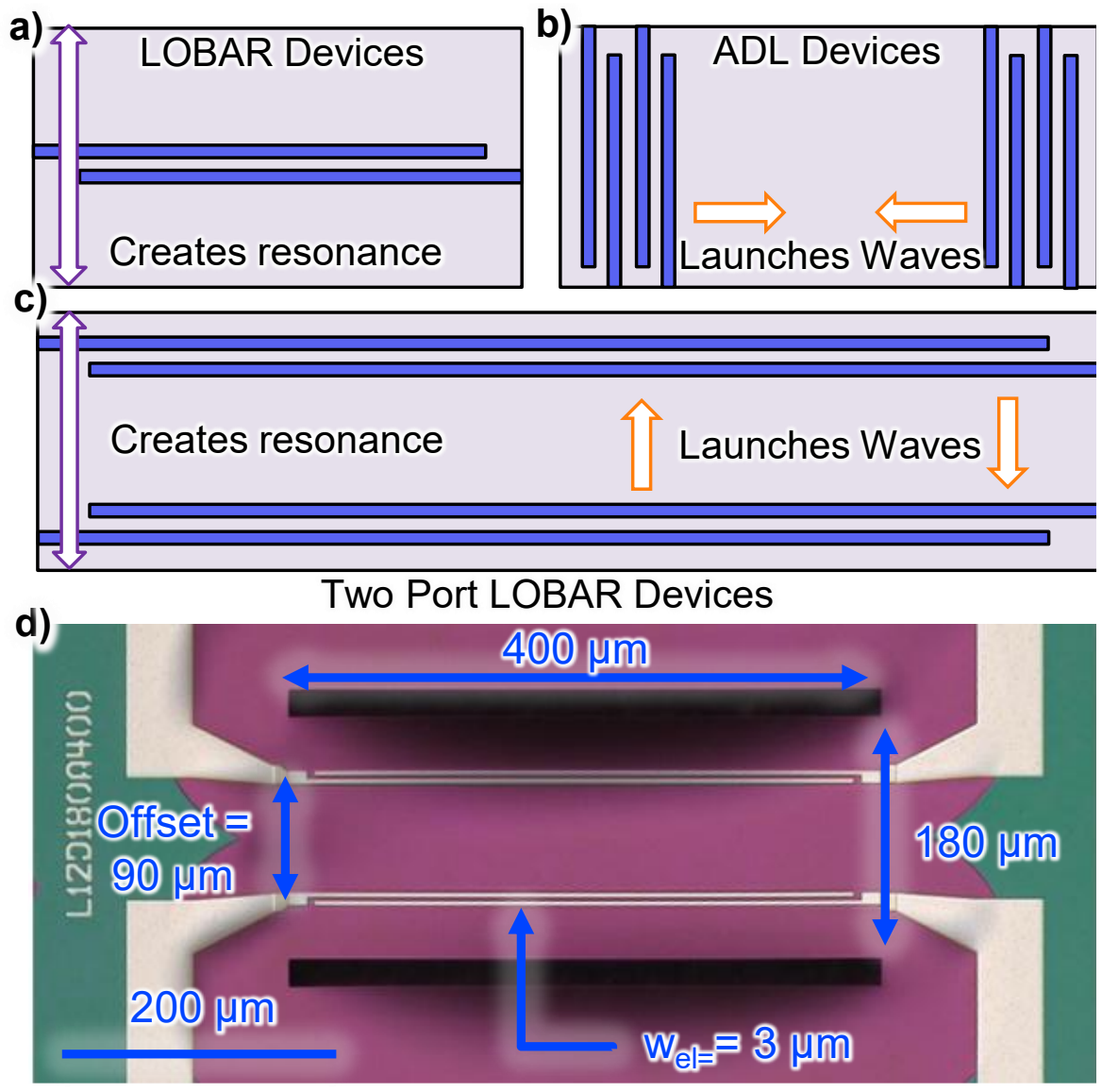

*Figure 2: (a) layout for typical LOBAR resonator (b) layout for acoustic delay lines (c) layout we use here, which combines the two behaviors, and (d) optical image of device with parameters.*

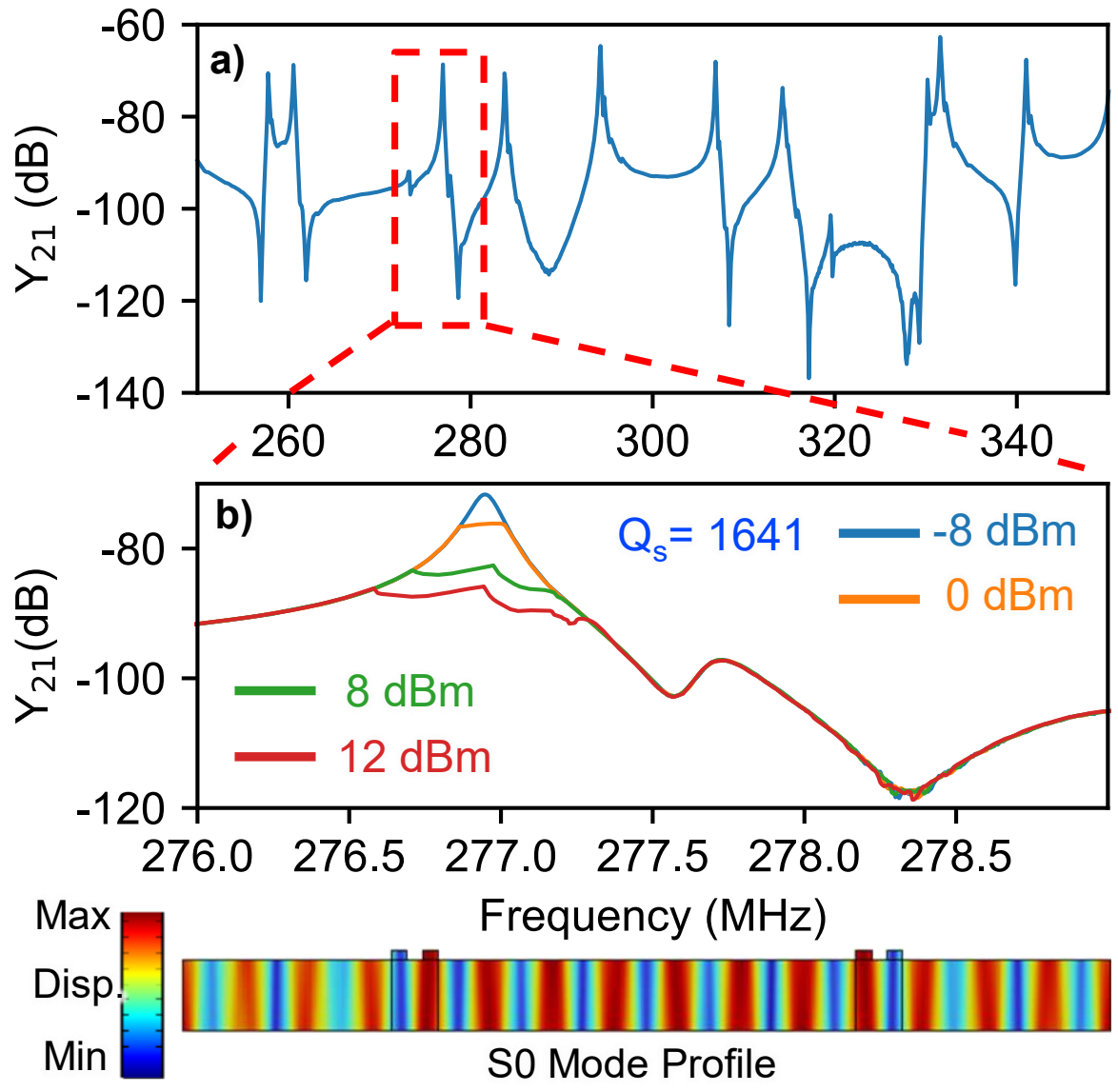

*Figure 3: (a) wider admittance profile of device showing acoustic delay lines and resonator characteristics (b) zoom in on resonance of interest with different powers, showing COMSOL simulation of mode below.*

## FABRICATION AND METHODOLOGY

We use a lateral overtone bulk acoustic resonator (LOBAR) structure similar to our previous demonstrations [13] of mechanical frequency combs. These devices leverage the high coupling of lithium niobate (LN) to couple into many modes with just a single set of electrodes [14]. Different from previous demonstrations, these devices contain two ports and an offset of the electrodes to couple into as many modes as possible. The result of such a device is a combination of an acoustic resonator and acoustic delay lines, seen in Fig. 2 (a) and (b). The device will resonate off the two etch windows but also deliver energy from one point to another (Fig. 1(c)). Optical images are additionally shown in Fig. 2 (d), where we highlight our use of λ = 12 µm as our lateral coupling wavelength and 180 µm. The devices are fabricated in 36Y LN at an in-plane angle of 22 degrees such that we couple into SH0 and S0 modes at the same time. Increasing the number of modes increases the probability that two lower-frequency modes satisfy the frequency sum requirement for parametric down-conversion [15]. The fabrication consists of deposition of 50 nm of aluminum via e-beam evaporation, etching of lithium niobate via argon gas physical ion milling, and release of devices with underside etching using $XeF_2$.

The resulting admittance profile is shown in Fig. 3(a), with a zoom-in on the resonance of interest in Fig. 3(b). It can be seen that we have resonance in the frequency range of 100 MHz to 1.5 GHz, with delay-line-like characteristics sitting on top of our resonances. Our resonances are the fundamental symmetric mode (S0) and fundamental shear horizontal mode (SH0), but we will focus specifically on the S0 mode resonance at 277 MHz for our measurements. Increasing power measurements show that our device shows significant Duffing bending of the resonance for powers above 0 dBm, originating from thermal nonlinearities [16]. Because this nonlinearity shifts the resonance as the drive power increases, the drive frequency and power must be chosen together when generating combs, which motivates the power tuning discussed below.

Mechanical frequency combs in said device are measured from 200 MHz all the way up to 900 MHz, but in order to stay in the range of our LDV system, we focus on our lower frequency 277 MHz mode. Figure 4 shows our electrical readout scheme, whereby we drive our device at a single frequency at high power using a signal generator. The electrical measurements are performed by contacting the device with RF probes, with the device driven at 276.046 MHz and 3 dBm. Because our device is a two-port device, we can avoid the use of couplers and use the second port as the output, fed to a signal analyzer (where the electrical signal from the input port is completely electrically isolated from the output. Fig. 4 (a)-(d) show the wideband response and the zoomed-in view of each of the driven and parametric tones. As expected, the driven tone is much stronger than the parametric resonances and gives rise to comb lines of Δf = 40.8 kHz. In addition to the comb tones, we observe a number of other tones in the spectrum, some of which also exhibit comb lines of their own. None of these tones are present at small drive powers, where the device is driven linearly, indicating that they arise from nonlinear behavior. When we measure with an LDV system, we hope to see a signal at this frequency; however, no tone exists in the electrical signal, as the motion of the device at this frequency range does not couple into our electrode geometry. Additionally, the spectrum analyzer noise in this frequency range is too large to see any small-strength signals that may be coupled in any way.

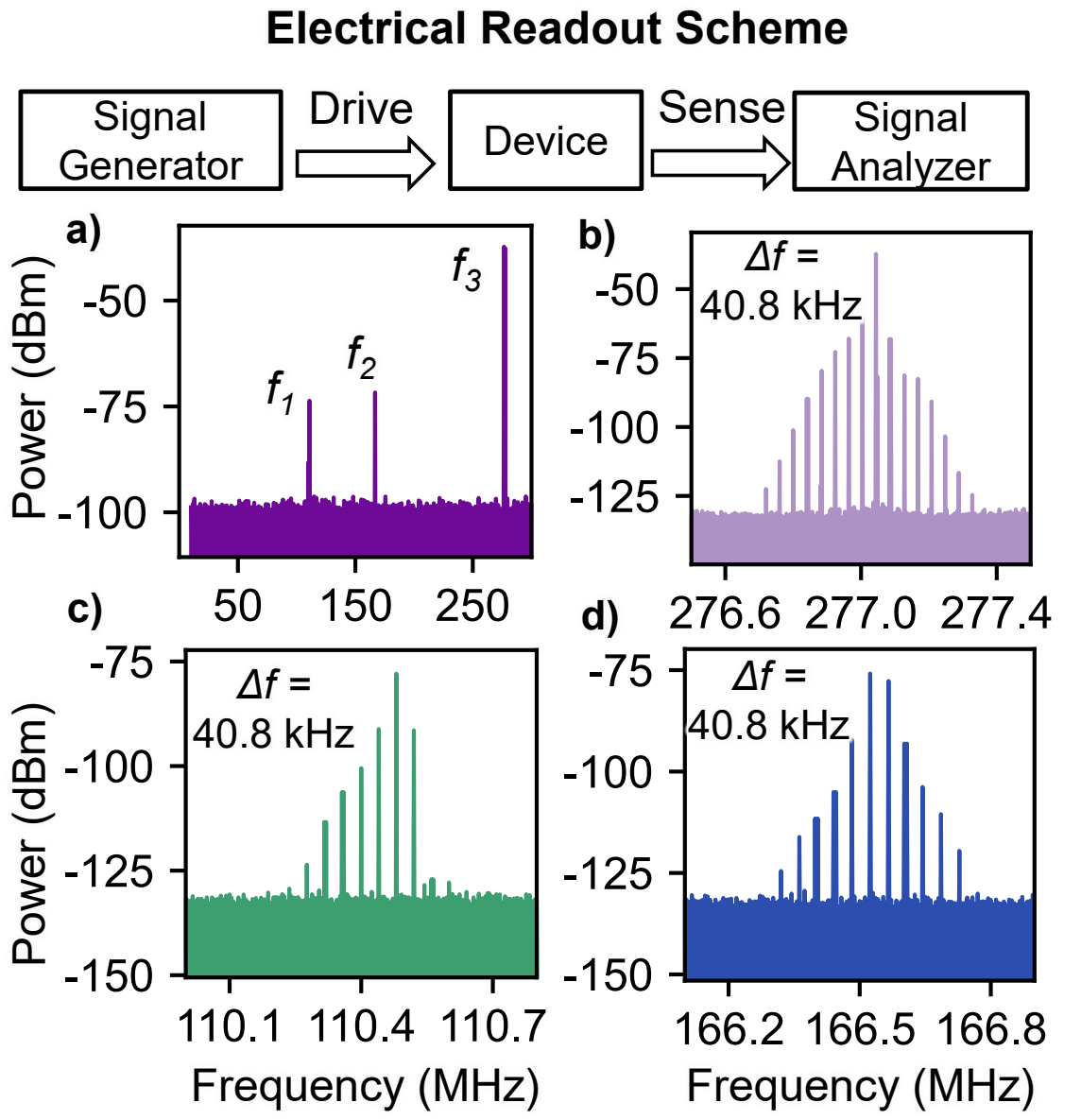

*Figure 4: (a) wide spectrum of electrical measurements, and (b)-(d) zoomed in measurements of each tone showing the mechanical comb and equal com spacing among the three tones.*

We move to our measurements using the LDV system to look at the low-frequency spectrum of the mechanical motion of the device. For these measurements, the device is wire-bonded so that it can be driven electrically while its out-of-plane motion is read out optically by the LDV. For the LDV measurements, the device is driven at 276.96 MHz and 15 dBm. Initial measurements of the device in the upper frequency range reveal the expected comb lines at each of the tones we saw in the electrical domain (Fig. 5). Two notable differences occur for the LDV measurements. Firstly, the higher frequency tones, instead of being stronger, are now weaker than the two lower parametric tones. This likely stems from the fact that the out-of-plane motion of the modes doesn't necessarily correspond to the strength of the electrical signal, as the modes are not primarily engineered to resonate out of plane. The second difference is the different comb spacing of Δf = 21 kHz. This difference likely comes from the introduced wire-bonding inductance giving a different solution for the comb when driven and measured for LDV. As seen in [13], different comb solutions of different spacing are easy to achieve via power tuning.

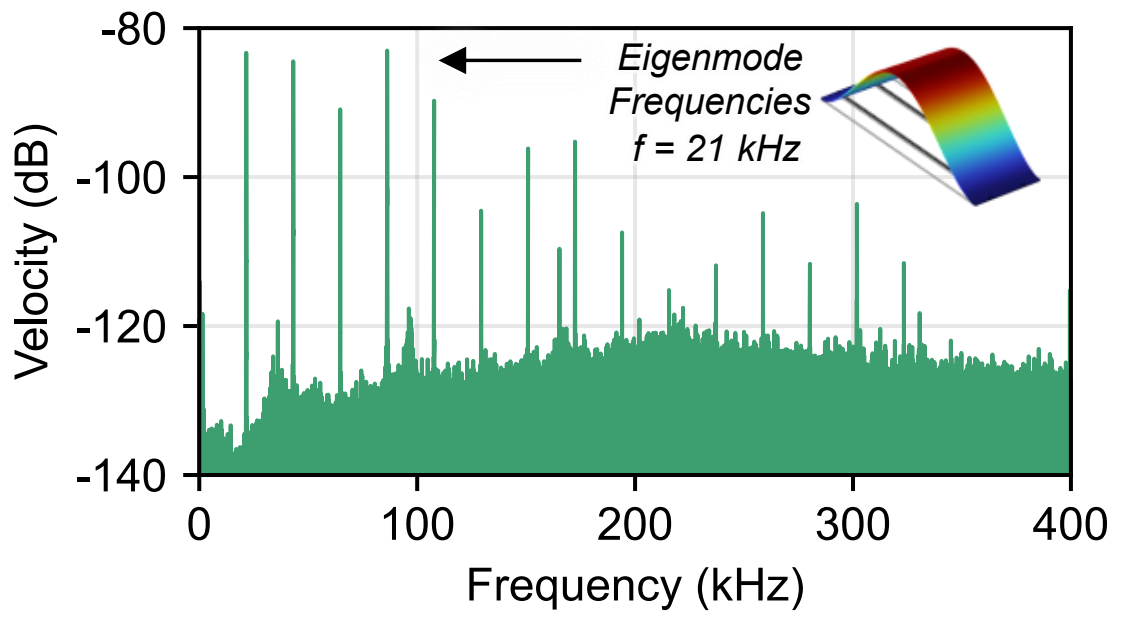

*Figure 6: output LDV spectrum showing nonlinear excited tones at the baseband from driving at high frequency.*

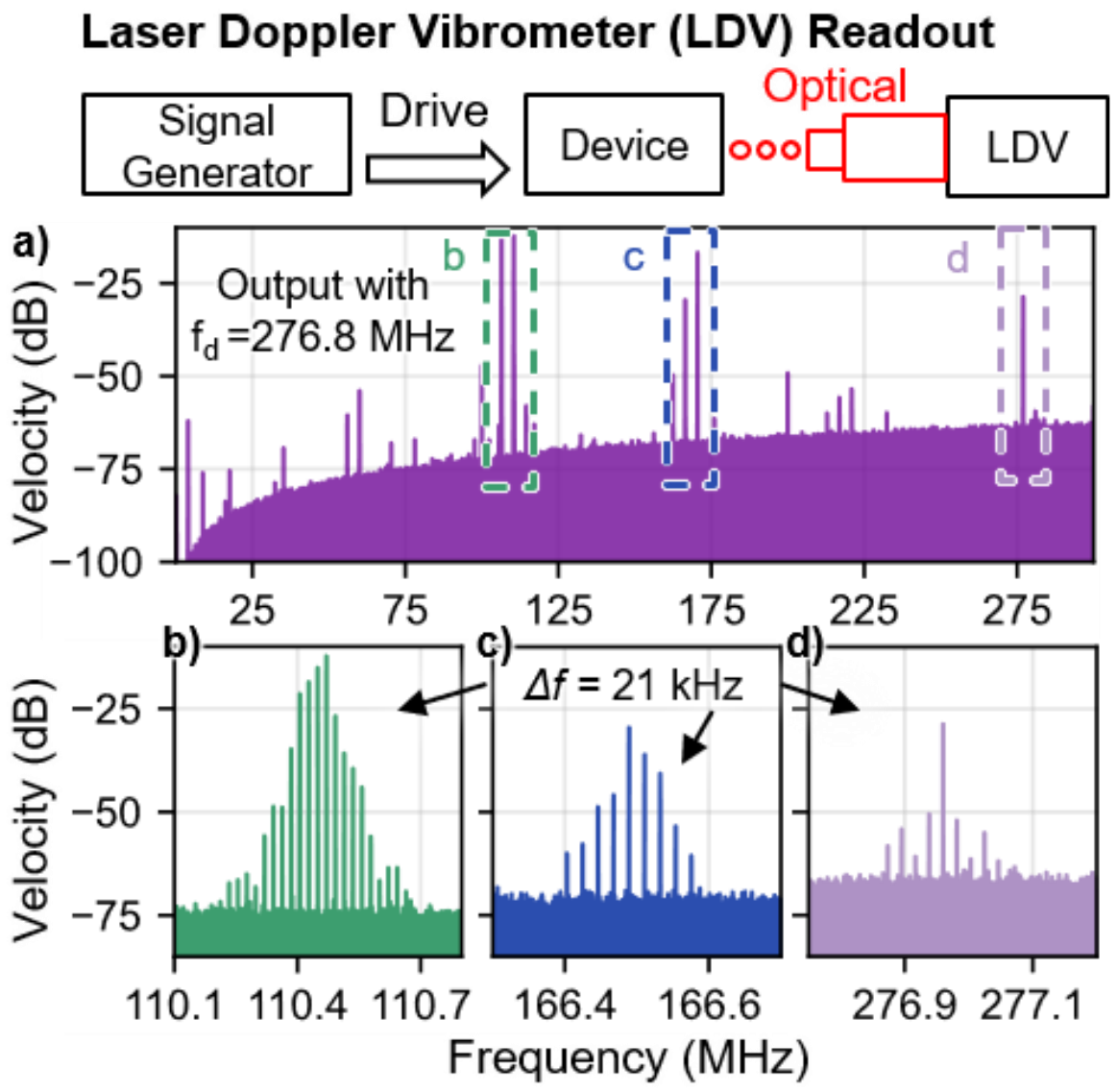

*Figure 5: (a) wide spectrum of LDV measurements showing a number of tones not shown in the electrical measurements (b)-(d) similar zoom in measurements of each comb device.*

We finally move on to the lower frequency spectrum of the devices in Fig. 6. We can see that the spectrum shows a tone at the exact frequency corresponding to the frequency spacing, along with higher-order overtones up to 300 kHz. This frequency corresponds exactly to that of the fundamental out-of-plane eigenmode of the device from COMSOL simulations. The spectrum additionally looks very similar to that of the photodetector outputs of optical frequency combs [17]. In the optical case, the photodetector converts the comb spacing into a low-frequency tone; here, the out-of-plane motion of the device plays the same role, with the 21 kHz tone and its overtones appearing at the comb spacing and its multiples. Relative to the 277 MHz drive, the 21 kHz tone represents a frequency division of roughly 13,000, or more than four orders of magnitude. These nonlinear measurements drive at a single frequency and have the desired quality that we can see all the frequencies at the same time. The downside to these nonlinear measurements is that we are unable to image these tones while they are being driven nonlinearly, and thus we cannot confirm if these tones correspond to resonance. Ideally, we will soon perform linear measurements of the high- and low-frequency response to see if these tones correspond to resonances, and whether this impacts the comb performance.

# CONCLUSION

In this work, we demonstrated mechanical frequency combs in the MHz regime for excitation of fundamental low-frequency tones in the kHz regime, corresponding to a conversion of energy over 4 orders of magnitude. This shows that the comb spacing generated by mechanical nonlinearities can be accessed directly as a low-frequency mechanical signal, analogous to the photodetector output of an optical frequency comb. While these tones are weaker than the upper-frequency tones, it is believed that strategically using different drive frequencies and powers to place the low-frequency tone on a resonance of the device will enhance the strength of said tones at low frequency. Future work will include linear measurements of the high- and low-frequency response to determine whether these tones correspond to resonances, as well as a study of how drive frequency and power can be used to optimize the low-frequency tone. Such devices could offer a compact, chip-scale route to bridging high-frequency mechanical signals with low-frequency electronics, with potential applications in frequency references and timing.

## ACKNOWLEDGEMENTS

This work was supported in part by the NASA Space Technology Graduate Research Opportunity (NSTGRO) grant No. 80NSSC24K1369, and the Defense Advanced Research Projects Agency Higher-Order Composite Resonators for Extra Resilience (HORCREX) program No. HR0011-25-9-0159. The authors appreciate the helpful discussions with Dr. Sunil Bhave, Dr. Andrea Alu, Dr. David Burghoff, Dr. Neal Hall, and Dr. Meng Kang. Any opinions, findings, conclusions, or recommendations expressed in this material are those of the author(s) and do not necessarily reflect the views of the Defense Advanced Research Projects Agency (DARPA).

## CONTACT

*Ian Anderson, email: ianderson@utexas.edu